\documentclass[prb,aps,graphics,epsfig,twocolumn,floats,superscriptaddress]{revtex4}
\usepackage{graphicx}
\usepackage{amssymb}
\usepackage{amsmath}
\usepackage{color}
\usepackage{cancel}
\usepackage{pifont}

\usepackage{setspace} 

\begin{document}

\title{Hysteresis and training effect in electric control of spin current in Pt/Y$_3$Fe$_5$O$_{12}$ heterostructures}

\author{Y. D. Sun$^{1,}$\footnotemark[2], Lei Wang$^{2,}$\footnotemark[2], Lili Lang$^{4}$, Ke Xia$^{3,}$\footnotemark[1], and S. M. Zhou}
\email{kexia@bnu.edu.cn}
\email{shiming@tongji.edu.cn}

\affiliation{Shanghai Key Laboratory of Special Artificial Microstructure Materials and Technology and Pohl Institute of Solid State Physics and School of Physics Science and Engineering, Tongji University, Shanghai 200092, China\\
$^{2}$Center for Spintronics and Quantum Systems, State Key Laboratory for Mechanical Behavior of Materials, Xi'an Jiaotong University, No.28 Xianning West Road Xi'an, Shanxi, 710049, China\\
$^{3}$School of Physics, Southeast University, Nanjing 211189, China\\
$^{4}$State Key Laboratory of Functional Materials for informatics, Shanghai Institute of Microsystem and information Technology, Chinese Academy of Sciences, Shanghai, 200050, China}

\date{\today}

\begin{abstract}
We have reported on the hysteresis and training effect of spin current in Pt/Y$_3$Fe$_5$O$_{12}$ heterostructures during subsequent cycles of ionic liquid gate voltage $V_g$. The inverse spin Hall effect voltage in spin pumping and spin Hall magnetoresistance exhibit diode-like behaviors in the first half cycle of $V_g$ and also show hysteresis in the first cycle of $V_g$. Both the diode-like behavior and the hysteresis become weak and even vanish in the second cycle of $V_g$ due to the training effect.
The above experimental results can be well explained by the screening charge doping model, in which the charge and the local magnetic moment are asymmetrically distributed in the Pt layer.
The applicability of this model is further confirmed by measurements of anisotropic magnetoresistance and ferromagnetic resonance.
The diode-like behavior is attributed to interplay between the asymmetrically distributed local magnetic moment and the spin current relaxation in the Pt layer. The hysteresis and the training effect arise from the irreversible interaction between the oxidation and reduction of Pt atoms and the evolution of the surface morphology at the ionic liquid/Pt interface under electric gating. This work provides new insights to improve the functional performance of electrically controlled spin current devices.

\end{abstract}

\maketitle
\footnotetext[2]{These authors contributed equally to this work.}

\section* {Introduction}

Due to the nonvolatile nature of spintronic devices, spin memory and logic have become important candidates for the development of next-generation chips, bringing spintronics back into the focus of academia and industrial research. The functional performance of spintronic devices depends on multiple parameters, such as the spin diffusion length (SDL) and spin Hall angle (SHA) in the heavy metals and the spin mixing conductance at the heavy metal/ferromagnet interface~\cite{Hoffman2013,Sinova2015}. Many intriguing spin current phenomena have emerged, such as spin Hall magnetoresistance (SMR)~\cite{PhysRevB.87.224401}, Hanle magnetoresistance (HMR)~\cite{PhysRevLett.116.016603}, and spin pumping~\cite{PhysRevLett.97.216603,PhysRevLett.88.117601}.

With growing demand for low energy consumption, the nondissipative feature of ionic liquid (IL) gating
has attracted significant attention in controlling magnetism~\cite{PhysRevB.99.224416,songc2015,PhysRevApplied.12.034005,Liang2018,PhysRevLett.111.216803}.
In particular, IL gating has shown great ability in tuning magnetic and spin current properties in Pt/Y$_3$Fe$_5$O$_{12}$ (YIG) ~\cite{Guan2018,dushenko2018tunable,yanss2019}. The ideal spin current system, embodying strong spin orbital coupling in Pt and a low magnetic damping factor in ferrimagnetic insulator YIG, has played a central role in spin current devices~\cite{Hoffman2013,Sinova2015}.
Classically, Dushenko \emph{et al.}~\cite{dushenko2018tunable} observed that the inverse spin Hall effect (ISHE) voltage in spin pumping experiments
remains unchanged for the gate voltage $V_g<0$, but it sharply decreases for $V_g>0$, displaying diode-like behavior
in the electric control of the spin current. The electric response of the spin current for $V_g>0$ was argued to arise from Fermi energy shift in the entire Pt layer.
Meanwhile, Guan \emph{et al.}~\cite{Guan2018} observed diode-like behavior in the electric response of a ferromagnetic resonance (FMR) magnetic field and inferred that the magnetic properties of the Pt layer remain unchanged when $V_g<0$.
In contrast, Wang \emph{et al.}~\cite{yanss2019} found that the ISHE voltage changes as a linear function of $V_g$ for both positive and negative $V_g$.

At present, the IL gating effect in metals is attributed to two individual effects, the Fermi level shift and the charge accumulation around the interfaces due to the screening effect~\cite{PhysRevB.99.224416}; the latter is denoted as the screening charge doping model (SCDM) in this work.
The key difference between these two effects lies in the spatial distribution of the local magnetic moment (LMM).
In the Fermi level shift model in Pt~\cite{dushenko2018tunable}, the Fermi energy in the entire Pt layer is assumed to shift toward lower energies when $V_g>0$, and the density of states (DOS) at the Fermi energy is increased because the DOS of Pt peaks slightly below the Fermi energy~\cite{kubler2017theory}. Consequently, the Stoner criterion is satisfied, and Pt atoms in the entire Pt layer are spin polarized. For $V_g<0$, the entire Pt layer is paramagnetic. In the SCDM, in contrast, the positive-charge doped Pt atoms in the Debye screening layer
around the Pt/YIG and IL/Pt interfaces can be spin polarized~\cite{PhysRevB.99.224416} to positive and negative $V_g$, respectively.

\begin{figure}[htbp]
	\centering
	\includegraphics[width=8cm]{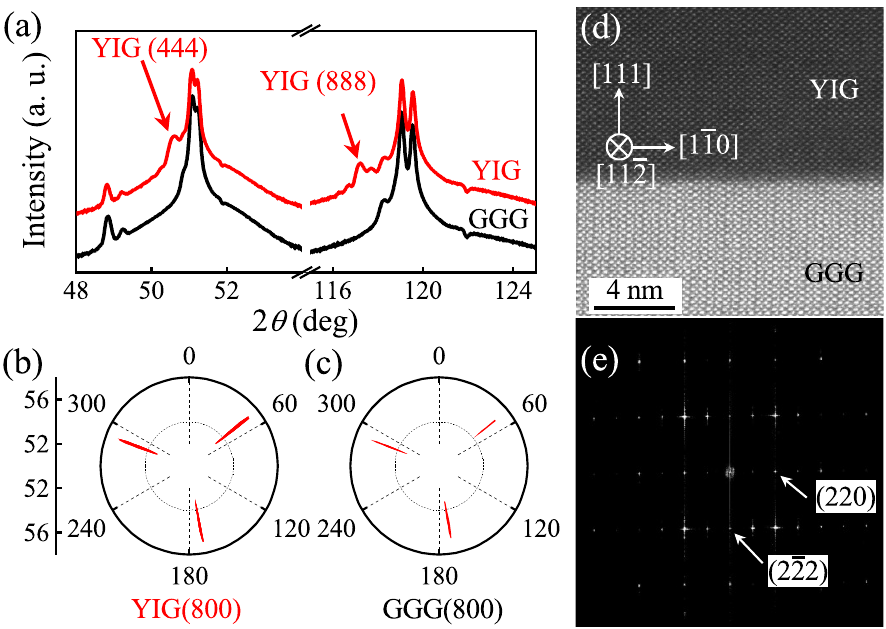}
\caption{(a) XRD $\theta$-2$\theta$ patterns of the GGG substrate (black) and the YIG film (40 nm thick) epitaxially grown on the GGG (444) substrate (red). X-ray pole figures of YIG (b) and GGG (c) along the [800] orientation. Atomic-resolution scanning transmission electron microscopy image along the [11$\overline{2}$] orientation (d) and selected area electron diffraction pattern (e) of YIG/GGG.}
	\label{Fig1}
\end{figure}

In this work, we report on the hysteresis and training effect of ISHE voltage and SMR in the Pt/YIG heterostructure during subsequent cycles of $V_g$. In the first half cycle of $V_g$, the ISHE voltage and the SMR remain unchanged for $V_g<0$ and decrease for $V_g>0$, exhibiting diode-like behavior. They also show hysteresis in the first cycle of $V_g$.
In the second cycle, the hysteresis and the diode-like behavior become weak and even vanish, exhibiting a training effect. Evidently, the electric control of the spin current is mainly determined by the charge spatial distribution in the Pt layer.
First-principles calculations and measurements of anisotropic magnetoresistance (AMR)~\cite{ChienMPE2012} and FMR confirmed the validity of the SCDM, in which the spatial distribution of LMM in the Pt layer is asymmetric between the Pt/YIG and IL/Pt interfaces.
Finally, the diode-like behavior, hysteresis, and training effect in the electric control of the spin current are well explained in terms of the SCDM.

\section* {Sample fabrication and measurements}
Pt (\emph{d})/YIG ($t_{\rm{YIG}}$) heterostructures were fabricated with \emph{d}=3.0 nm and $t_{\rm{YIG}}$=40 nm and 50 nm. YIG single-crystal films were epitaxially fabricated onto (444)-oriented Gd$_3$Ga$_5$O$_{12}$ (GGG) substrates via pulsed laser deposition at 625$^{\circ}$C. The base pressure was lower than $2\times 10^{-6}$ Pa, and the O$_2$ pressure was 3.0 Pa during deposition. The deposition rate of the YIG layer was 0.33 $\rm{\mathring A}$/s. To improve the film quality, the samples were post-annealed at 810$^{\circ}$C for 4 hours under an O$_2$ pressure of $5\times10^4$ Pa. A Pt layer was fabricated on the YIG layer at ambient temperature by DC magnetron sputtering. The base pressure was better than $5\times 10^{-6}$ Pa, the Ar pressure was 0.5 Pa during deposition, and the deposition rate of the Pt layer was 0.70 $\rm{\mathring A}$/s.
We used X-ray reflectivity (XRR) and X-ray diffraction (XRD) to characterize
the layer thickness and high crystallinity of the samples, respectively, with a Bruker D8 diffractometer with Cu K$\alpha1$ ($\lambda_1$ =0.154056 nm) and K$\alpha2$ ($\lambda_2$ =0.154439 nm). A typical XRD pattern of the YIG layer is presented in Fig.~\ref{Fig1}(a). The peaks near 2$\theta$ = 50.58$^{\circ}$ and 117.22$^{\circ}$ correspond to the (444) and (888) orientations in the YIG layer, respectively, while the double peaks of GGG (444) correspond to K$\alpha1$ and K$\alpha2$.
The pole figures of the YIG layer and the GGG substrate in Figs.~\ref{Fig1}(b) and ~\ref{Fig1}(c) demonstrate the epitaxial growth of the YIG (444) layer on the GGG (444) substrate. The cross-sectional scanning transmission electron microscopy and the selected area electron diffraction pattern in Figs.~\ref{Fig1}(d) and ~\ref{Fig1}(e) also show the epitaxial relationship between the YIG film and the GGG substrate.\\
\indent To impose a significant electric field on the film, the IL [DEME]$^+$[TFSI]$^-$ was employed as the dielectric material. A small droplet of IL was placed on the film device.
Before measurements of spin pumping, SMR, and AMR, a gate voltage was applied on the device and then maintained at room temperature for 30 min. to enable the formation of a stable electric field on the film surface. Cooling the device below 182 K lead to freezing of the IL, resulting in a fixed anion and cation distribution (and therefore electric field).

\begin{figure}[tbp]
	\centering
	\includegraphics[width=8cm]{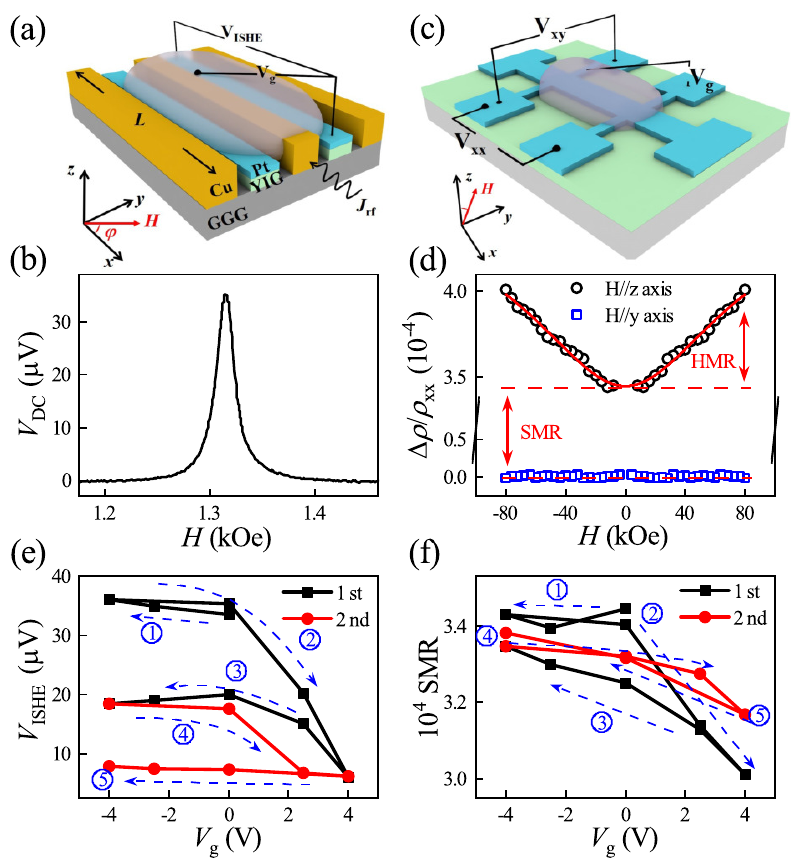}
\caption{Diode-like behavior, hysteresis, and training effect of ISHE voltage and SMR. Schematic illustration of spin pumping (a) and SMR and HMR (c) measurements.
For Pt(3.0 nm)/YIG(40 nm), typical spin pumping spectrum (b), magnetic field dependence of $\Delta\rho/\rho_{\rm{xx}}$ in SMR and HMR (d), evolutions of
ISHE voltage (e) and SMR (f) during cycles of the gate voltage. In (b, e), $\varphi$ = 90 degrees and $f$ = 6 GHz. In (b, d), $V_g=0$.
In (d), solid lines refer to fitted results according to Eqs.\ref{pauli4} and \ref{pauli5}. In (e,f), arrows and symbols \ding{192}-\ding{176} indicate the gate voltage sweep. Here, all measurements were performed at $T=100$ K. }
	\label{Fig2}
\end{figure}

\indent We then used the spin pumping technique to characterize the ISHE voltage, Gilbert damping parameter, and gyromagnetic ratio ($\gamma$) of the Pt/YIG heterostructure~\cite{PhysRevLett.97.216603,PhysRevLett.88.117601,MaAEM2016}, as schematically shown in Fig.~\ref{Fig2}(a).
By feeding a microwave signal into the coplanar waveguide, a RF magnetic field perpendicular to the sample strip was induced to excite the magnetization precession in the YIG layer.
When sweeping the in-plane magnetic field $H$, the DC voltage detected by a lock-in amplifier peaks at the resonance magnetic field $H_{\rm{RES}}$
at a specific microwave frequency $f$. As defined by the peak amplitude in Fig.~\ref{Fig2}(b), the ISHE voltage $V_{\rm{ISHE}}$ obeys the following equation~\cite{spin-pumping-2010}
\begin{equation}
V_{\rm{ISHE}}=-\frac{eL\omega G_{\rm{mix}}^{\rm{eff}}\theta_{\rm{SH}}\lambda_{sd}\rho_{\rm{xx}}}{2\pi d}\tanh(\frac{d}{2\lambda_{\rm{sd}}})\theta^2\sin\varphi,
\label{pauli1}
\hspace{0.0 cm}
\end{equation}
where $\theta_{\rm{SH}}$, $\lambda_{\rm{sd}}$, $d$, $\rho_{\rm{xx}}$, and $G_{\rm{mix}}^{\rm{eff}}$ are the SHA, SDL, thickness, sheet resistivity of the Pt layer, and real part of the effective spin mixing conductance (ESMC) at the Pt/YIG interface, respectively. Moreover, $\omega=2\pi f$, the stripe length $L=2.0$ mm, $\theta$ is the precessional angle of the YIG magnetization, and $\varphi$ refers to the angle between the external magnetic field and the $x$ axis, as shown in Fig.~\ref{Fig2}(a).

At a specific $f$, the central resonance magnetic field $H_{\rm{RES}}$ of the measured spin pumping spectrum can be fitted by a Lorentz function, and the dispersion for the Pt/YIG heterostructure can be described by the Kittel equation~\cite{Kittel-equation}
\begin{equation}
(\frac{\omega}{\gamma})^2=H_{\rm{RES}}(H_{\rm{RES}}+4\pi M_{\rm{eff}}),
 \label{pauli2}
\hspace{0.0 cm}
\end{equation}
where the effective demagnetization field is $4\pi M_{\rm{eff}}=4\pi M_s+H_K$, $M_{\rm{eff}}$ is the effective magnetization,
$M_s$ is the saturation magnetization, and $H_K$ is the magnetic anisotropic field. Using Eq.~\ref{pauli2}, the gyromagnetic ratio $\gamma(= -g|e|/2cm_e)$ is fitted, where $g$ is the Land\'{e} factor and $e$, $m_e$, and $c$ are the charge and mass of electrons and the speed of light in vacuum, respectively. With the full width at half maximum (FWHM) linewidth $\Delta H$ of the resonance as a function of $f$,
the Gilbert damping parameter $\alpha_{\rm{Pt/YIG}}$ in the Pt/YIG heterostructure can be achieved using the following
equation~\cite{PhysRevLett.107.066604}
\begin{equation}
\Delta H = \Delta H_0 + \frac{4\pi\alpha_{\rm {Pt/YIG}}}{\gamma}f,
\label{pauli3}
\hspace{0.0 cm}
\end{equation}
where $\Delta H_0$ is the broadening width induced by the inhomogeneity of the YIG layer. Moreover, FMR was employed to measure the Gilbert damping parameter $\alpha_{\rm {YIG}}$ of the YIG single-layer film~\cite{Kittel-equation}.
Then, the ESMC $G_{\rm{eff}}^{\rm{mix}}$ in the Pt/YIG heterostructure can be obtained through the following equation \cite{PhysRevLett.107.066604}
\begin{equation}
G_{\rm{eff}}^{\rm{mix}}=\frac{4\pi M_{s} t_{\rm {YIG}}} {g \mu_{B}} (\alpha_{\rm {Pt/YIG}}-\alpha_{\rm {YIG}}),
\label{pauli6}
\hspace{0.0 cm}
\end{equation}
where $\mu_B$ is the Bohr magneton and $t_{\rm {YIG}}$ is the YIG layer thickness.
In this work, the IL gating effect on the magnetic properties of the YIG layer is neglected~\cite{Zhao_2021}.

The SMR and AMR in the Pt/YIG heterostructure were measured by a standard four-point probe, as schematically shown in Fig.~\ref{Fig2}(c).
With the external magnetic field $H$ along the $y$ and $z$ axes, SMR and HMR are defined in Fig.~\ref{Fig2}(d)\cite{PhysRevB.87.224401,PhysRevLett.116.016603,xu2019gate,PhysRevB.100.064404}. SMR is described by the following equation~\cite{PhysRevB.87.224401}
\begin{equation}
(\Delta\rho/\rho_{\rm{xx}})_{\rm{1}}=-\frac{\theta_{\rm{SH}}^2\lambda_{\rm{sd}}}{d}\frac{\tanh^2(d/2\lambda_{\rm{sd}})}{1/(2\rho_{\rm{xx}}\lambda_{\rm{sd}}G_{\rm{eff}}^{\rm{mix}})+\coth(d/\lambda_{\rm{sd}})}.
\label{pauli4}
\hspace{0.0 cm}
\end{equation}
Due to the dephasing of electron spins caused by spin precession around the external magnetic field~\cite{PhysRevLett.116.016603}, HMR is rigorously described using the following equation~\cite{PhysRevLett.116.016603}
\begin{equation}
\begin{split}
&(\Delta\rho/\rho_{\rm{xx}})_{\rm{2}}=-Re\{\frac{\theta_{\rm{SH}}^2\Lambda_{\rm{sd}}}{d}\frac{\tanh^2(d/2\Lambda_{\rm{sd}})}{1/(2\rho_{\rm{xx}}\Lambda_{\rm{sd}}G_{\rm{mix}}^{\rm{eff}})+\coth(d/\Lambda_{ \rm{sd}})}\} \\
&-\{-\frac{\theta_{\rm{SH}}^2\lambda_{\rm{sd}}}{d}\frac{\tanh^2(d/2\lambda_{\rm{sd}})}{1/(2\rho_{\rm{xx}}\lambda_{\rm{sd}}G_{\rm{mix}}^{\rm{eff}})+\coth(d/\lambda_{\rm{sd}})}\},
\end{split}
\label{pauli5}
\hspace{0.0 cm}
\end{equation}
where the first and second terms on the right-hand side refer to SMRs at high and zero magnetic fields, respectively.
Considering spin precession around the external magnetic field $H$, the effective SDL obeys the following equation $1/\Lambda_{\rm{sd}}=\sqrt{1/\lambda_{\rm{sd}}^2+i/\lambda_m^2}$, where $\lambda_m=\sqrt{D\hbar/g\mu_BB}$, $\hbar$, $D$, and $B$ represent the reduced Planck constant, electron diffusion coefficient, and magnetic induction intensity, respectively.
When $\Lambda_{\rm{sd}}=\lambda_{\rm{sd}}$ at $H=0$, the HMR vanishes.
A new approach was proposed by V\'{e}lez \emph{et al.} to independently extract SDL and SHA~\cite{PhysRevLett.116.016603}, and it was improved later by Dai \emph{et al.}~\cite{PhysRevB.100.064404}
such that both parameters can be rigorously extracted. Being independent of SHA, the HMR/SMR ratio can be employed to extract SDL with the ESMC data.
Subsequently, SHA can be obtained with data of SMR, SDL and ESMC through Eq.~\ref{pauli4}. In measurements of the AMR and anomalous Hall-like effect (AHLE)~\cite{PhysRevB.92.060402}, the external magnetic field was rotated in the \emph{xz} plane and aligned along the \emph{z} axis.
\begin{figure}[tbp]
	\centering
	\includegraphics[width=8cm]{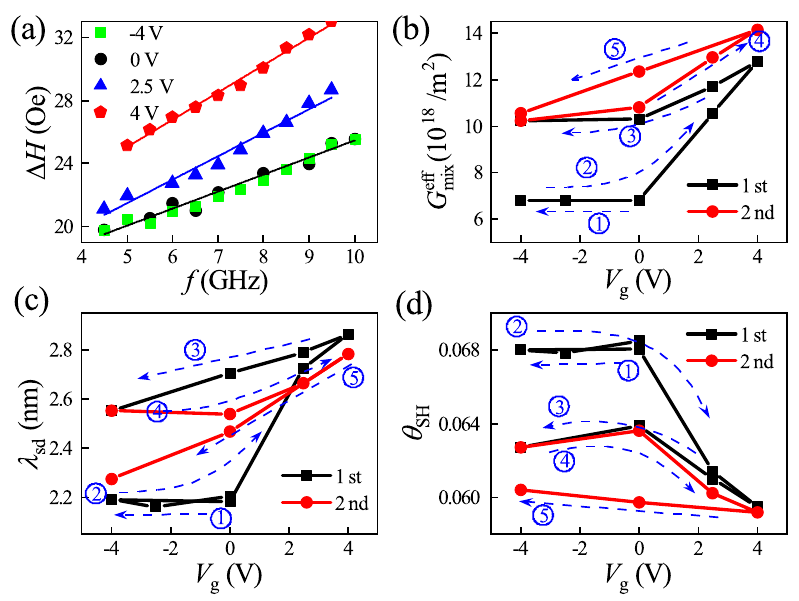}
\caption{Diode-like behavior, hysteresis, and training effect of ESMC, SDL, and SHA. For Pt(3.0 nm)/YIG(40 nm), the $\Delta H$ \emph{vs.} $f$ at different $V_g$ (a). ESMC $G_{\rm{mix}}^{\rm{eff}}$ (b), SDL $\lambda_{sd}$ (c), and SHA $\theta_{\rm{SH}}$ (d) \emph{vs.} $V_g$ in the first (black) and second (red) cycles. In (a), the inset numbers refer to the gate voltage, and solid lines refer to fitted results according to Eq.~\ref{pauli3}.
In (b, c, d), the arrows and symbols \ding{192}-\ding{176} indicate the gate voltage sweep. Here, all measurements were performed at 100 K.}
	\label{Fig3}
\end{figure}

\section* {First-principles calculations}
We then performed first-principles calculations using the Vienna ab initio simulation package (VASP) ~\cite{Kresse1993,Kresse1996} to assess the IL gating effect. The calculations were based on DFT and generalized gradient approximation (GGA) with an interpolation formula as given by Vosko, Wilk, and Nusair~\cite{Vosko1980} and a plane-wave basis set within the framework of the projector augmented wave (PAW) method~\cite{Blochl1994,Kresse1999}. The cutoff energy for the basis was 500 eV, and the convergence criterion for the electron density self-consistency cycles was 10$^{-6}$ eV. In the Brillouin zone, we sampled ($15\times15\times15$) k-point grids using the Monkhorst-Pack scheme~\cite{Monkhorst1976} to ensure that the results converged. The details of the first-principles calculations were
analytically described in a previous study~\cite{Guan2018}.

\section* {Results and discussion}
\emph{Diode-like behavior, hysteresis, and training effect in electric response of spin current in Pt/YIG}. {\textemdash}
As shown in Figs.~\ref{Fig2}(e) and ~\ref{Fig2}(f), the ISHE voltage and SMR exhibit a few distinguishing features. (1) In the first half cycle of $V_g$, the ISHE voltage and SMR remain almost unchanged when $V_g$ increases from -4.0 to 0 (V), while they decrease when $V_g$ further increases up to 4.0 V, demonstrating diode-like behavior~\cite{dushenko2018tunable}. It is noted that the ISHE voltage changes more sharply than SMR. (2) Both the ISHE voltage and SMR exhibit hysteresis when $V_g$ sweeps, indicating that the electric response of the spin current is irreversible. Since the ISHE voltage and SMR at $V_g=-4.0$ V are reduced after one cycle of $V_g$, the spin current is partially recovered. (3) In particular, the diode-like behavior and hysteresis of both the ISHE voltage and SMR become weaker or even vanish in the second cycle of $V_g$ compared to those of the first cycle, exhibiting a training effect in the electric response of the spin current.

Furthermore, Eqs.~\ref{pauli1} and ~\ref{pauli4} suggest that the ISHE voltage and SMR strongly depend on SDL, ESMC, and SHA~\cite{Hoffman2013,Sinova2015}. To
unravel the origin of the results in Fig.~\ref{Fig2}, it is essential to extract the above three spin current parameters
at various $V_g$.
As shown in Fig.~\ref{Fig3}(a),
the FWHM linewidth $\Delta H$ in the Pt/YIG heterostructure scales as a linear function of the microwave frequency $f$, and
$\alpha_{\rm {Pt/YIG}}$ in Eq.~\ref{pauli3} is found to change with $V_g$.
ESMC, SDL, and SHA in Figs.~\ref{Fig3}(b-d) exhibit diode-like behavior, hysteresis, and training effect during subsequent cycles of $V_g$, similar to ISHE voltage and SMR. In particular, when $V_g$ increases from 0 to 4.0 V in the first half cycle of $V_g$, SDL and ESMC increase, whereas SHA decreases.\\

\begin{figure}[htbp]
	\centering
	\includegraphics[width=8cm]{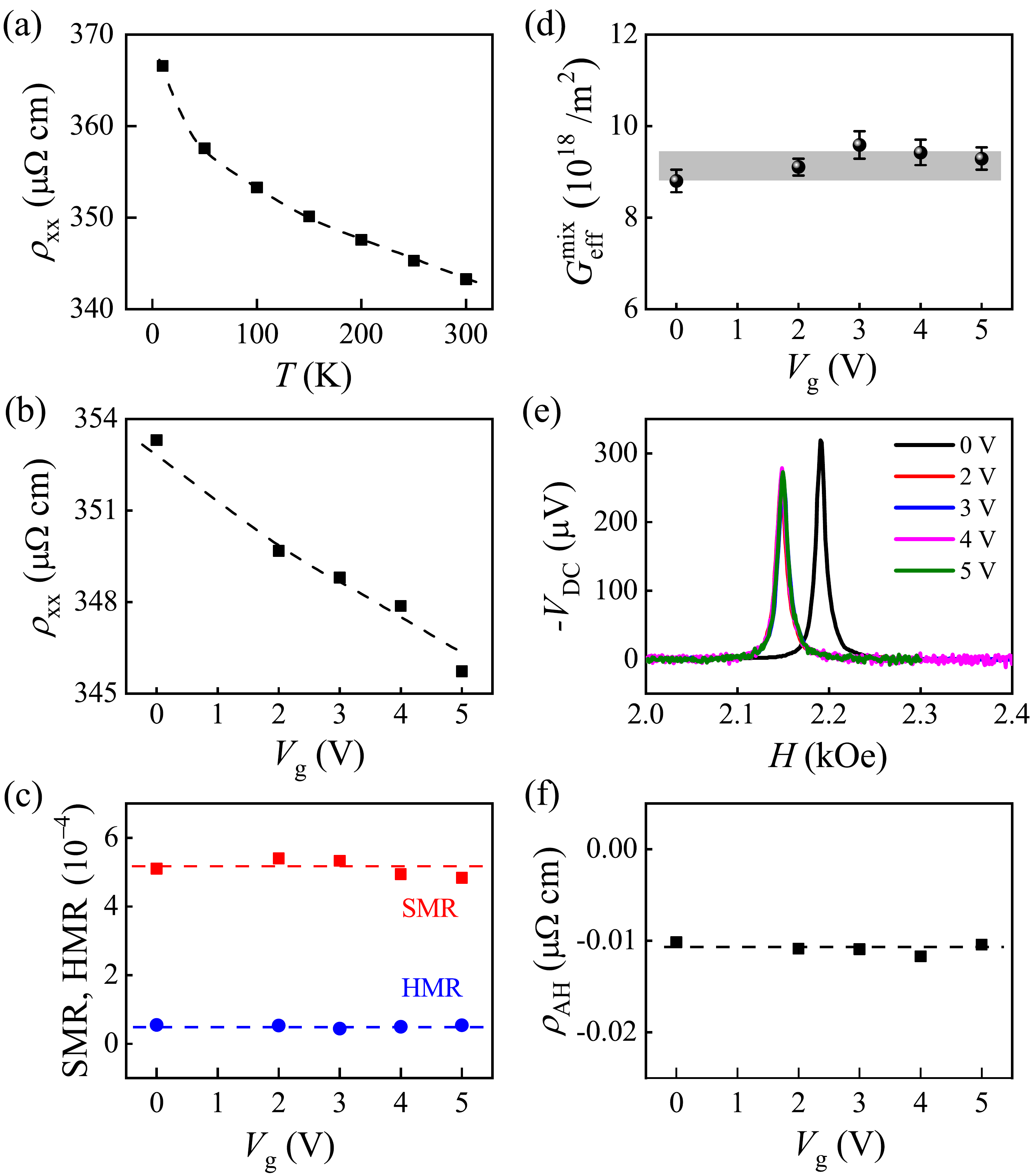}
\caption{Electric control of the spin current in $\beta$-W (3.0 nm)/YIG (40 nm) heterostructures. The sheet resistivity $\rho_{\rm{xx}}$ versus temperature at $V_g=0$ (a), $V_g$ dependencies of $\rho_{\rm{xx}}$ (b), SMR and HMR (c), ESMC (d), typical spin pumping spectra at different $V_g$ (e), $V_g$ dependence of anomalous Hall-like resistivity $\rho_{\rm{AH}}$ (f). In (e), $\varphi=90$ degrees and the rf frequency $f$=9.0 GHz. Dashed lines in (a, b, c, f) serve as a guide to the eye. Measurements in (b, c, d, e, f) were performed at $T=100$ K.}
	\label{Fig4}
\end{figure}

For comparison, we also investigated the electric response of the spin current in the $\beta$-W (3.0 nm)/YIG (40 nm) heterostructure.
The $\beta$ phase of the W layer is identified by the nonmetallic behavior~\cite{beta-W}, where $d\rho_{\rm{xx}}/dT<0$ in Fig.~\ref{Fig4}(a).
In our experiments, the electric response of the spin current in $\beta$-W/YIG is found to almost vanish, although the sheet resistivity decreases for $V_g>0$, as shown in Figs.~\ref{Fig4}(b-f). Moreover, since the SHA and sheet resistivity of the $\beta$-W layer are much larger than those of the Pt layer~\cite{pai2012-Beta-W-SHA,Hoffman2013}, the ISHE voltage in the $\beta$-W/YIG heterostructure is approximately one order of magnitude larger than that of the Pt/YIG heterostructure when the input microwave power is identical and $V_g$ is zero in the two experiments, as shown in Figs.~\ref{Fig2}(b) and~\ref{Fig4}(e).
Furthermore, although $M_{\rm{eff}}$ remains unchanged, the gyromagnetic ratio $\gamma$ is 16.4 and 16.6 (GHz/kOe) for $V_g=0$ and $V_g>0$, respectively, leading to a shift of the resonance field for $V_g>0$ away from that of $V_g=0$, as shown in Fig.~\ref{Fig4}(e).\\

\begin{figure}[tbp]
	\centering
	\includegraphics[width=8cm]{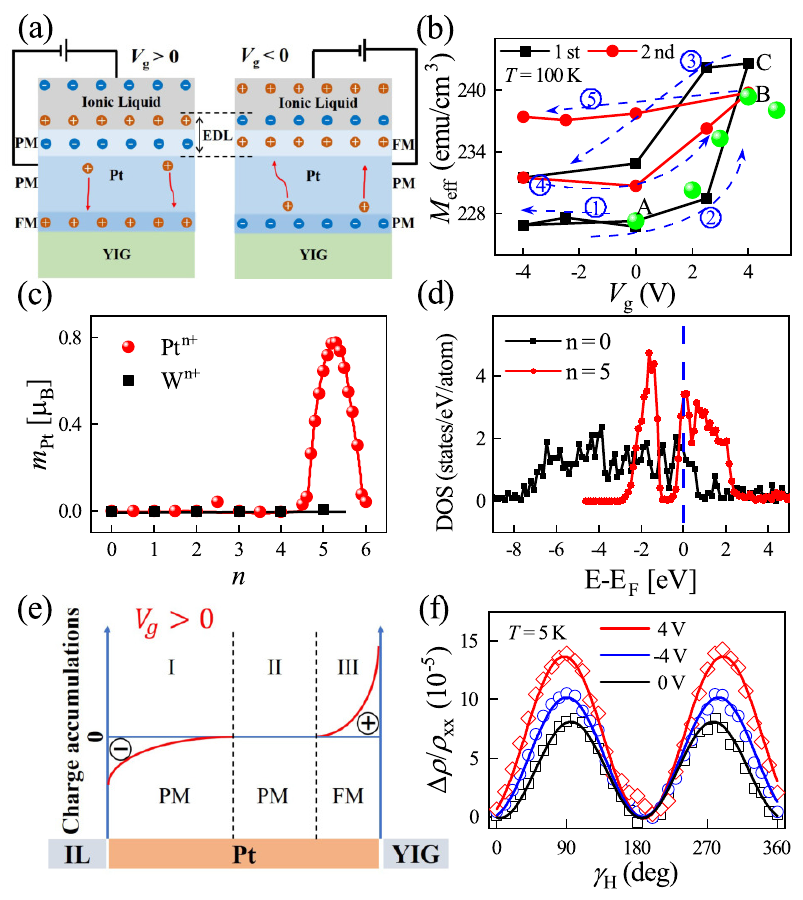}
\caption{Asymmetric spatial distributions of charge and LMM in Pt. For Pt(3.0 nm)/YIG(40 nm), sketch for charge spatial distribution for $V_g>0$ and $V_g<0$ (a), $M_{\rm{eff}}$ versus $V_g$ at $T=100$ K (b), for Pt$^{n+}$ atoms with $V_g>0$,
calculated magnetic moment $m_{\rm{Pt}}$ \emph{vs.} $n$ (c) and DOS \emph{vs.} $E-E_F$ with $n=0$ and 5 (d), sketch for the charge asymmetric spatial distribution in the Pt layer for $V_g>0$ (e), angular dependent $\Delta\rho/\rho_{\rm{xx}}$ at $T=5$ K with the magnetic field rotating in the $xz$ plane and $V_g$ changing in the order of 0, -4.0, and 4.0 (V) (f). In (a, e), FM and PM stand for ferromagnetic and paramagnetic states, respectively.
In (b), the data (green solid circles) of Pt(3.0 nm)/YIG(50 nm) in the first half cycle of the IL gate voltage are also given for comparison, and the arrows and symbols \ding{192}-\ding{176} indicate the gate voltage sweep, $\Delta M_{\rm{eff}}=M_{\rm{eff}}(B)-M_{\rm{eff}}(A)=12.1$ emu/cm$^3$
for $t_{\rm{YIG}}=50$ nm and $\Delta M_{\rm{eff}}=M_{\rm{eff}}(C)-M_{\rm{eff}}(A)=16.0$ emu/cm$^3$ for $t_{\rm{YIG}}=40$ nm.
In (c), the data of the $\beta$-W/YIG heterostructure (black solid squares) are also given.
In (e), regions I and III are negatively and positively charged, respectively, and region II is electrically neutral.
Solid lines in (c, f) serve as a guide to the eye and refer to fitted results, respectively. In (f), $\Delta\rho=\rho_{\rm{xx}}\sin^2{\gamma_H}$ and $\gamma_H$ refer to the angle between $H$ and the $z$ axis in AMR measurements. }
	\label{Fig5}
\end{figure}

\emph{Asymmetric spatial distributions of charge and LMM in Pt/YIG. }{\textemdash} Since the results in Fig.~\ref{Fig3} indicate that SHA, SDL, and ESMC are multivalued functions of $V_g$, the electric control of the spin current is not caused by the Fermi level shift but by the charge spatial distribution in the Pt layer, which is characterized by the results of the LMM in Fig.~\ref{Fig5}.
The effective magnetization $M_{\rm{eff}}$ shows diode-like behavior
in the first half cycle of $V_g$~\cite{Guan2018,zhao2017}, as shown in Fig.~\ref{Fig5}(b). A negligibly small $H_K$ in the magnetically soft YIG layer, as defined in Eq.~\ref{pauli2}, results in $M_{\rm{eff}}\simeq\frac{m_{\rm{tot}}}{t_{\rm{YIG}}\Delta S}$, where $\Delta S$ and
$m_{\rm{tot}}$ are the area and total magnetic moment of the sample, respectively. Defined as the increment of $M_{\rm{eff}}$ when $V_g$ changes from 0 to 4.0 V in the first half cycle of $V_g$, $\Delta M_{\rm{eff}}$ is 16.0 and 12.1 (emu/cm$^3$) for $t_{\rm{YIG}}=40$ and 50 (nm), respectively, as shown in Fig.~\ref{Fig5}(b).

Since the magnetic moment of the YIG layer is independent of $V_g$~\cite{Zhao_2021} and assuming that there is no magnetic proximity effect in the Pt/YIG heterostructure~\cite{ChienMPE2012}, the diode-like behavior
of $M_{\rm{eff}}$ in the first half cycle can be easily characterized by the Fermi level shift model as the DOS of Pt peaks slightly below the Fermi energy~\cite{kubler2017theory}. In the SCDM, in contrast,
The IL gating establishes an electric double layer (EDL) at the IL/Pt interface, as shown in Fig.~\ref{Fig5}(a).
For $V_g>0$, positive and negative charges accumulate on the IL and Pt sides of the EDL, respectively.
Accordingly, positive charges in region I are shifted toward region III due to the Debye-screening effect, as marked in Figs.~\ref{Fig5}(a) and ~\ref{Fig5}(e), and Pt atoms in region III become positively charged, Pt$^{n+}$. The results of first-principles calculations in Fig.~\ref{Fig5}(c) show that the magnetic moment of Pt$^{n+}$ emerges for $4.5\leq n<6$. For bulk Pt without IL gating, the product of the DOS at the Fermi energy with the Stoner parameter is estimated to be approximately 0.6~\cite{MacDonald1981}.
As shown in Fig.~\ref{Fig5}(d), the DOS at the Fermi energy is enhanced by a factor of 2.0 when $n$=5.3.
Consequently, the product of the DOS at the Fermi energy with the Stoner parameter will be approximately 1.2 if the Stoner parameter remains the same,
and the Stoner criterion for the appearance of ferromagnetism is satisfied~\cite{PhysRevB.99.224416}.
Therefore, Pt atoms in region III contribute to $M_{\rm{eff}}$. Meanwhile, the Pt atoms in
Regions I and II are paramagnetic. Conversely, the Pt atoms are ferromagnetic in region I and paramagnetic in region III for $V_g<0$. Since the LMM in region I is not exchange-coupled to the YIG magnetization due to paramagnetic regions II and III acting as a spacer, the measured $M_{\rm{eff}}$ remains unchanged when $V_g<0$, as shown in Fig.~\ref{Fig5}(b). Since both models work, the FMR is not an efficient method to evaluate the LMM spatial distribution in the Pt layer.

Different from FMR measurements, \emph{all} Pt atoms in regions I, II, and III contribute to AMR. Thus, we performed AMR measurements at $T=5$ K to further identify the physical origin of the LMM. Figure~\ref{Fig5}(f) shows that AMR is enhanced not only when $V_g>0$ but also when $V_g<0$, which \emph{contradicts} the first mechanism dictating that there should be no AMR when $V_g<0$.
Because AMR enhancement with highly doped Pt$^{n+}$ occurs for either positive or negative $V_g$, the LMM in region I also contributes to AMR when $V_g<0$~\cite{ChienMPE2012}.
Therefore, SCDM can reconcile both the diode-like behavior of
$M_{\rm{eff}}$ in Fig.~\ref{Fig5}(b) and the AMR results in Fig.~\ref{Fig5}(f).
Finally, it is noted that the LMM consists of two parts. In addition to the part induced by charge doping under IL gating, the second part comes as a result of the magnetic proximity effect~\cite{ChienMPE2012,Liang2016-MPE}, as demonstrated by a nonzero AMR at $V_g=0$ in Fig.~\ref{Fig5}(f). Since the second part is independent of $V_g$, we focus only on the first (and the LMM only refers to this first part).

To quantitatively confirm the mechanism of the LMM in the SCDM, it is necessary to compare the experimental results and first-principles calculations. Considering that magnetization reaches a maximum at approximately $V_g=4.0$ V in experiments and approximately $n=5.3$ in first-principles calculations in Figs.~\ref{Fig5} (b) and ~\ref{Fig5}(c), it is reasonable to compare the data. To avoid the complex relation between $V_g$ and $n$, we assume that all Pt$^{n+}$ atoms in region III have the same valence state $n$.
Since $M_{\rm{eff}}$ enhancement derives exclusively from the Pt atoms in region III,
the estimated enhancement from the first-principles calculations will be $\Delta M_{\rm{eff}}=\frac{LMM}{t_{\rm{YIG}}\Delta S}=\frac{M_{\rm{Pt}}l_D}{t_{\rm{YIG}}}$, where $l_D$ is the screening length of region III and $M_{\rm{Pt}}$ represents the saturation magnetization of Pt atoms in this region. Thus, when the atomic magnetic moment of Pt$^{n+}$ is $m_{\rm{Pt}}=0.78~\mu_{B}$ at $n=5.3$, $l_D$ is estimated to be 1.34 nm and 1.26 nm with $\Delta M_{\rm{eff}}=16.0$ and 12.1 (emu/cm$^3$) for $t_{\rm{YIG}}=40$ and 50 (nm), respectively.

Calculations also show that ferromagnetism cannot be produced in $\beta$-W by any charge doping, as shown in Fig.~\ref{Fig5}(c).
Different magnetic properties in Pt/YIG and $\beta$-W/YIG heterostructures
can be understood as follows~\cite{kubler2017theory}: There are two implicit conditions for the generation of ferromagnetism: a) unpaired spin-resolved orbitals and b) strong localization of the orbitals. Although unpaired spin-resolved orbitals exist in bulk Pt 5d$^9$6s$^1$, ferromagnetism fails to occur because $d$ electrons are weakly localized.
When introducing positive charge doping and $n\geq4.5$, the $d$ electrons become strongly localized,
as demonstrated by the enhanced DOS near the Fermi energy for $n=5.3$ in Fig.~\ref{Fig5}(d). Accordingly, ferromagnetism occurs in Pt$^{n+}$ atoms with $4.5\leq n<6.0$. However, positive charge doping cannot induce any magnetic moment in $\beta$-W because of the much weaker localization of \emph{d} orbitals in $\beta$-W 5d$^4$6s$^2$ compared with that of Pt 5d$^9$6s$^1$.
Furthermore, because magnetism greatly affects spin transport, the AMR, ISHE voltage, and ESMC in the $\beta$-W/YIG heterostructure should be more insensitive to positive charge doping than those in the Pt/YIG heterostructure with Pt acting nearly as a ferromagnet.
\\

\emph{Mechanism of diode-like behavior in the electric control of SDL, SHA, and ESMC in Pt/YIG}{\textemdash} The SDL results in Fig.~\ref{Fig3}(c) are ascribed to an interplay between the asymmetrically distributed LMM and the spin current relaxation process, as analyzed below. Our experiments revealed that the electronic diffusion coefficient $D$ monotonically decreases from $5.8\times 10^{-6}$ to $3.7\times 10^{-6}$ (m$^2$/s) when $V_g$ increases up to 4.0 V in the first half cycle. Since $D=v_F^2\tau_e$, the electron momentum relaxation time $\tau_e$ is also expected to decrease with increasing $V_g$~\cite{Lang_2018}. Additionally, because $\lambda_{\rm{sd}}=\sqrt{D\tau_s}$, the increase in SDL with $V_g$ in Fig.~\ref{Fig3}(c) indicates that the spin current relaxation time $\tau_s$ becomes longer at a larger $V_g$.
Opposite variation trends of $\tau_s$ and $\tau_e$ suggest that the spin current relaxation process in the present Pt/YIG heterostructure is dominated by the D'yakonov-Perel' mechanism~\cite{PhysRevB.100.064404,PhysRevB.98.224424,dyakonov1972spin}, unlike the results of Dushenko \emph{et al.}~\cite{dushenko2018tunable}. In this perspective, $\tau_s$ is mainly determined by the $\tau_e$ of Pt atoms in region III.
With the presence of LMM in the same region when $V_g>0$, $\tau_e$ is reduced significantly, yet it changes little in the absence of LMM when $V_g<0$. Therefore, the diode-like behavior of $\lambda_{\rm{sd}}$ in Fig.~\ref{Fig3}(c) can be simply characterized.

The diode-like behavior of SHA in the first half cycle of $V_g$ in Fig.~\ref{Fig3}(d) can also be understood in a similar way.
Owing to the D'yakonov-Perel' mechanism of the spin current relaxation~\cite{PhysRevB.100.064404,PhysRevB.98.224424,dyakonov1972spin}, $\theta_{\rm{SH}}(III)$ plays a major role in the measured SHA of the Pt layer, in contrast to $\theta_{\rm{SH}}(I)$ and $\theta_{\rm{SH}}(II)$, which play minor roles. Accordingly, we observe $\theta_{SH}\approx\theta_{\rm{SH}}(III)$.
With the presence of the LMM in region III for $V_g>0$, $\theta_{\rm{SH}}(III)$ is significantly reduced due to the spin splitting of the chemical potential~\cite{zhang2015MPESHA,Guo2014}, whereas it changes little with the absence of the LMM for $V_g<0$. Consequently, the measured SHA exhibits diode-like behavior in the first half cycle of $V_g$ in Fig.~\ref{Fig3}(d). Finally,
since the LMM in region III can contribute to the Gilbert damping parameter of the Pt/YIG heterostructure~\cite{Swindells2021}, the latter exhibits diode-like behavior in the first half cycle of $V_g$.\\

\emph{Mechanisms of hysteresis and training effect in electric control of spin current in Pt/YIG. }{\textemdash} Since the charge $Q$ in the EDL is at the heart of the SCDM, it is of vital importance to gain deep insight into the evolution of $Q$ with $V_g$.
The magnitude of $Q(=CV_g)$ is proportional to the capacitance $C$ of the capacitor between two electrodes,
and the latter depends on the surface morphology of the IL and the oxidation and reduction of Pt atoms at the IL/Pt interface~\cite{Walsh2014-oxide-layer,voroshylova2020hysteresis}.
Due to these irreversible and nonequilibrium evolutions, $C$ is not constant and rather exhibits hysteresis and training effects during subsequent cycles of $V_g$~\cite{voroshylova2020hysteresis, Walsh2014-oxide-layer,Bhatt2006,Druschler2010-EC-window}.
Therefore, $Q$ is expected to have similar behaviors when $V_g$ is swept.

\begin{figure}[tbp]
	\centering
	\includegraphics[width=8cm]{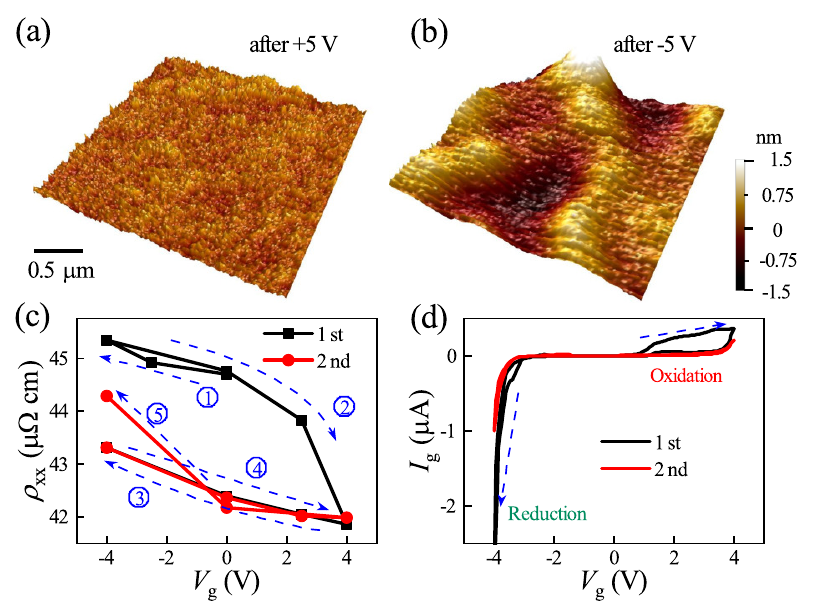}
\caption{Oxidation and reduction of Pt atoms at the IL/Pt interface. Atomic force microscopy images of the Pt surface after applying positive (a) and negative $V_g$ (b) with a scanned area size of 2$\times2$ $\mu m^2$. The sheet resistivity at $T$=100 K during subsequent cycles of the gate voltage (c). Cyclic voltammetry with a sweep rate of 100 mV/s for Pt(3.0 nm)/YIG(40 nm) (d). In (c), the arrows and symbols \ding{192}-\ding{176} indicate the variation of the gate voltage. Measurements in (a, b, d) were performed at room temperature. }
	\label{Fig6}
\end{figure}

The above analysis is verified by the experimental results in Fig.~\ref{Fig6}. First,
atomic force microscopy measurements show that the surface root mean square (RMS) roughness $R_q$ of the
Pt layer at the IL/Pt interface remains unchanged for $V_g>0$, whereas it increases from 0.143 nm to 0.443 nm when $V_g$ changes from 0 to -4.0 V, as shown in Figs.~\ref{Fig6}(a) and ~\ref{Fig6}(b).
Second, the asymmetric distribution of the LMM results in a decrease in the sheet resistivity of the Pt layer.
This is because the resistance of ferromagnetic region III is increased, whereas that of paramagnetic region I is decreased when $V_g>0$. Similarly, the application of negative $V_g$ also results in a decrease in the sheet resistivity. As a result, the sheet resistivity is expected to change nonmonotonically as a function of $V_g$, with a maximum at $V_g=0$. Apparently, the monotonic variation in the sheet resistivity in Fig.~\ref{Fig6}(c) is consistent with the larger surface roughness of the Pt layer for $V_g<0$ in Fig.~\ref{Fig6}(b).
The incompletely reversible process between oxidation and reduction of Pt atoms and the evolution of the surface morphology at the IL/Pt interface are suggested to play major and minor roles in the surface degradation of the electrode (Pt) under electric gating~\cite{zhao2017,wang2021-interfacial-restructuring}, respectively.

The incompletely reversible process between oxidation and reduction of Pt atoms permits the dynamic resistance to alter both the effective voltage on the Pt/YIG heterostructure and the subsequent leakage current~\cite{Walsh2014-oxide-layer,Bhatt2006}. Figure~\ref{Fig6}(d) shows the hysteresis and training effect of the peak currents near $V_g=4.0$ and -4.0 (V) in the cyclic voltammetry, where these peak positions correspond to the oxidation and reduction of the Pt electrode.
Therefore, the experimental results in Fig.~\ref{Fig6} are consistent with the previous argument describing the incompletely reversible process between oxidation and reduction of Pt atoms and the evolution of the surface morphology at the IL/Pt interface under electric gating~\cite{Walsh2014-oxide-layer,voroshylova2020hysteresis}.
Moreover, due to the law of charge conservation, the charge $Q$ in the EDL is also known to exhibit hysteresis and training effects in cyclic voltammetry.
Since the LMM depends on $Q$, the electric control of the spin current in Figs.~\ref{Fig2}-\ref{Fig3} can be easily captured using the SCDM.
Again, the electrochemical stability window in the IL/Pt system is characterized by cyclic voltammetry to be 8.0 V~\cite{Druschler2010-EC-window}.

Reversible electric control of spin current is technologically immensely important in applications of spintronic devices and can be implemented by two approaches. First, since the leakage current in the cyclic voltammetry changes reversibly in the second cycle of $V_g$ in Fig.~\ref{Fig6}(d), a reversible electric response of the spin current is also expected after many cycles. Spin current measurements in one cycle of $V_g$,
at \emph{low temperatures}, requires tens of hours, which is much longer than the time scale of several minutes in cyclic voltammetry at \emph{room temperature}. Second, when the sweep range of $V_g$ is much smaller than the electrochemical window of the IL/electrode (Pt),
the leakage current and thus the spin current also change reversibly with $V_g$.
Since the electric response of the spin current cannot be maximized in this configuration, it is essential to find ideal ILs with a large electrochemical stability window for significant and reversible modulation of the spin current by IL gating~\cite{dushenko2018tunable}.

This work helps to solve the discrepancy in the electric response of Pt/YIG among different research groups. The reversible electric control of the spin current in Ref.~\onlinecite{yanss2019} may arise from two possible sources. First, when $n\leq4.5$ due to a small charge $Q$ in the EDL or a large Debye screening length in the Pt layer, LMM fails to occur, and the electric response of the spin current in the Pt/YIG heterostructure is thus determined by the Fermi level shift. Second, the sweep range of $V_g$ is smaller than the electrochemical window of IL/Pt. In these two cases, the electric response of the spin current is reversible.
Moreover, the diode-like behavior of the ISHE voltage in Ref.~\onlinecite{dushenko2018tunable} may not come as a result of a Fermi level shift but instead from a combination of the asymmetrically distributed LMM and the spin current relaxation process. Furthermore, since the LMM in region I and the YIG magnetization are separated by paramagnetic regions II and III acting as spacers, as schematically shown in Fig.~\ref{Fig5}(a), FMR cannot probe the LMM near the IL/Pt interface for $V_g<0$. Hence, a conclusion on the magnetic properties of the Pt layer for $V_g<0$ could not be drawn only from FMR measurements~\cite{Guan2018}. Finally, the opposite variation trends of $\theta_{SH}$ and
$M_{\rm{eff}}$ with $V_g$ in Figs.~\ref{Fig3}(d) and~\ref{Fig5}(b) \emph{directly} confirm the magnetic proximity effect on SHA of Pt~\cite{zhang2015MPESHA}, and explain inconsistencies of SHA in Pt reported in literature~\cite{Hoffman2013,Sinova2015,Taoscienceadvances2018}, considering the run-to-run variation of magnetic proximity effect in individually fabricated samples~\cite{Liang2016-MPE}.\\

\emph{In summary}. {\textemdash} We have experimentally demonstrated the hysteresis and training effect during electric control of spin current in the Pt/YIG heterostructure through subsequent cycles of $V_g$. SHA, SDL, and ESMC all manifest diode-like behavior, hysteresis, and training effects during $V_g$ cycling, leading to similar behaviors in ISHE voltage and SMR. Therefore, the electric response of the spin current in the Pt layer cannot be explained by the Fermi level shift but by the SCDM. In this model, the charge and the LMM are suggested to be asymmetrically distributed in
the Pt layer for either positive or negative $V_g$, which is confirmed by AMR and FMR measurements. In combination with the spin current relaxation process, the asymmetrically distributed LMM generates diode-like behavior in the electric control of the spin current. Meanwhile,
The incompletely reversible process between the oxidation and reduction of Pt atoms and the evolution of the surface morphology at the IL/Pt interface under electric gating result in the hysteresis and training effect in the electric control of the spin current. Moreover, opposite variation trends of SHA and LMM with $V_g$ further confirm the magnetic proximity effect on SHA in Pt. The relaxation process of the spin current is dominated by the D'yakonov-Perel' mechanism.
The present work will ignite a surge of interest in electric control of spin current, which has potential applications in energy-efficient spintronic devices.\\

\emph{Acknowledgements}.{\textemdash}
This study was supported by National Natural Science Foundation of China (Grants No. 11874283 and 11734004).

\end{document}